# Lightweight Call-Graph Construction for Multilingual Software Analysis

Anne Marie Bogar[1], Damian M. Lyons[1], and David Baird[2]
[1]*Department of Computer and Information Science, Fordham University, New York NY U.S.A.*
[2]*Bloomberg L.P., New York NY U.S.A.*
*{abogar, dlyons}@fordham.edu, dbaird16@bloomberg.net*

Abstract: Analysis of multilingual codebases is a topic of increasing importance. In prior work, we have proposed the MLSA (*MultiLingual Software Analysis)* architecture, an approach to the lightweight analysis of multilingual codebases, and have shown how it can be used to address the challenge of constructing a single call graph from multilingual software with mutual calls. This paper addresses the challenge of constructing monolingual call graphs in a lightweight manner (consistent with the objective of MLSA) which nonetheless yields sufficient information for resolving language interoperability calls. A novel approach is proposed which leverages information from a compiler-generated AST to provide the quality of call graph necessary, while the program itself is written using an Island Grammar that parses the AST providing the lightweight aspect necessary. Performance results are presented for a C/C++ implementation of the approach, PAIGE (*Parsing AST using Island Grammar Call Graph Emitter*) showing that despite its lightweight nature, it outperforms *Doxgen*, is robust to changes in the (*Clang*) AST, and is not restricted to C/C++.

## 1 INTRODUCTION

Large companies and software projects often face the challenge that, either for historical or for functionality reasons, they include code in different languages (Mushtak and Rasool, 2015). While software engineering support for single-language development has matured, support for multilingual development is not as strong (Mayer, 2017). In prior work (Lyons, Bogar, and Baird, 2017) we presented the MLSA (*MultiLingual Software Analysis*) architecture – a lightweight architecture concept for static analysis of multilingual software, and we showed that one important software engineering tool for dealing with multilingual code is call graph analysis. We showed how it was possible to stitch multilingual programs into a single call graph which could then be analyzed for safety, quality and other software engineering metrics.

However, constructing a call graph, particularly for languages with objects and first-class functions, can be time-consuming: (Ali and Lhotak, 2012) cites an example of a 30 second processing time to produce a call graph for a "Hello World" program in Java, due to the complications of dynamic dispatch.

Call graph generation is not a new topic and many algorithms exist (e.g., (Bacon and Sweeney, 1996) (Srivastava, 1992) (Dean, Grove, and Chambers, 1995)). While software that creates call graphs already exist, such as *Eclipse* and *Doxygen*, these tools usually do not provide adequate information for multilingual analysis: handling pointers, first-class functions and classes, as well as providing argument value information (needed to disambiguate interoperability (Barrett, Kaplan, and Wileden, 1996) calls).

This paper proposes a novel approach to call graph generation, PAIGE (***P**arsing **A**ST using **I**sland **G**rammar Call Graph **E**mitter*) with the following characteristics:

- **Lightweight:** Leverage existing software to extract the required information from source code.
- **Modular:** Process the extracted information in a manner that is not bound to a specific language.
- **Static Analysis:** All potential calls are captured rather than the subset seen in any single run.

PAIGE operates on a textual Abstract Syntax Tree (AST) generated by the same monolingual compiler used to compile the code. PAIGE is written as an *Island Grammar* (Moonen, 2002) so the approach is not tied to a single language. This paper focuses on PAIGE for C/C++ (using the *Clang* AST) but the grammar has also been modified for Python and JavaScript.

The next section briefly introduces MLSA, for context. Next PAIGE is introduced, explaining the AST and Island Grammar and their use in call graph construction. Performance results are presented for a 100-program comparison of PAIGE and *Doxygen*, showing PAIGE's ability to leverage a full AST for a superior result despite its lightweight, Island Grammar implementation. Results are also presented showing source language independence and robustness to AST changes.

## 2 THE MLSA ARCHITECTURE

Modern software development is increasingly multilingual: Developers might build a software project from different language components for functionality or style reasons, while at the same time companies maintain a commitment to language already used in company software, leading to an increasingly crowded and complex landscape of multilingual software development. Any solution that is narrowly focused on the existing state-of-the-art will find itself quickly outdated as new languages or interoperability APIs or language embeddings appear. For this reason, we propose the following design principles for a software architecture for *MultiLingual Software Analysis* - the MLSA architecture.

1. **Lightweight**: Computation is carried out by small programs, which we call *filters*, that communicate results with each other. A specific software analysis is built as a pipeline of these programs.

2. **Modular**: Filter programs accept input and produce output in a standard form, to allow modules to be substituted or added with minimum collateral damage.

3. **Open**: MLSA uses open-source software for monolingual processing and for display.

4. **Static Analysis**: MLSA uses static analysis as its principle approach

A generic example software analysis MLSA pipeline is illustrated in Figure 1. Pipelines are generally divided into three layers:

1. The initial filter programs consume a monolingual AST generated by an appropriate monolingual parser in the *monolingual layer* of the architecture.

2. The programs that consume the monolingual ASTs to process interoperability APIs must be also language specific; this happens in the *interoperability layer*.

3. In the final, *multilingual layer*, all the program data has been transformed to multilingual, and procedures in different languages can be related to one another for static analysis.

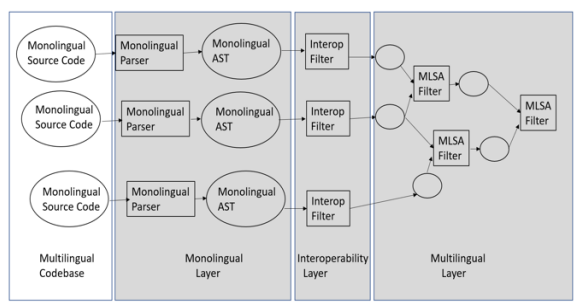

Figure 1: Example MLSA software analysis showing filter programs (squares) and their standardized input and output (circles) operating on a multilingual codebase (unshaded right-hand side)

In (Lyons, Bogar, and Baird, 2017) we presented an illustrative example of a Reaching Definitions Analysis (Nielson, Nielson, and Hankin, 2005) (RDA) pipeline for a C program.

- A monolingual AST filter `CASN` inputs the C AST and generates an output data file of variable assignment locations.
- A second filter `CRCF` generates the reverse control flow as a data file.
- A multilingual filter `RDA` takes the variable assignment location table and reverse control flow table as inputs, calculates reaching definitions for each variable, and writes those to a data file.

Each filter is a small standalone program, implementing a single analysis, illustrating the lightweight design principle. Filter programs can be written in different languages or can be shell scripts.

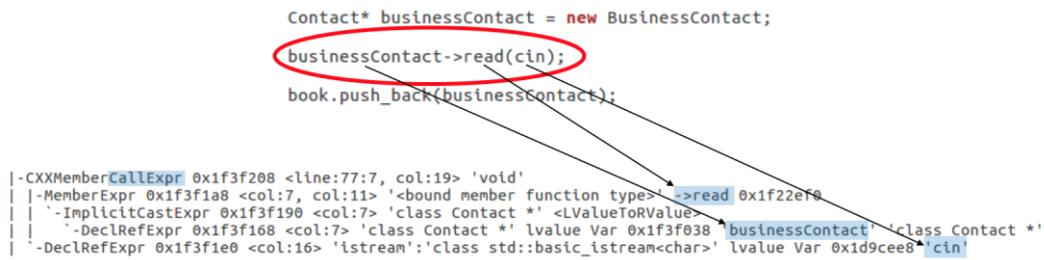

Figure 2: Clang C++ Abstract Syntax Tree, Example Procedure Call Node

The PAIGE program operates as a filter at the monolingual level of the architecture: It inputs source files (C/C++ in this paper) and generates procedure call information that can be later processed by interoperability and multilingual analysis filters. PAIGE can be a lightweight process because it relies on two concepts: an Abstract Syntax Tree generated by the monolingual compiler as input, and an Island Grammar to specify how the input is processed. We begin by reviewing these two concepts in context.

## 3 ISLANDS AND TREES

An *Abstract Syntax Tree (AST)* is a tree structure of the syntax of a program, in which each block – such as an if-statement, variable assignment, or function call – is a node with child nodes consisting of attributes of that block and other blocks inside that block. ASTs are generally used by compilers during semantic analysis when language-specific errors, such as type mismatch or out-of-scope variable usage, are caught. ASTs are also generally used by code analysis tools and IDEs such as *Eclipse*, as they present the code in a detailed manner that is easy to parse for specific types of blocks. MLSA uses the AST specifically generated by *Clang* for C/C++ programs (e.g., Figure 2); the Python ast module is used for Python and *SpiderMonkey* is used for JavaScript.

Figure 2 depicts how a function call in a program is expressed in the *Clang* AST. The function call node is identified by the keyword `CallExpr` or `CXXMemberCallExpr` for member functions. The name of the function, `read`, is a child node of the `CallExpr` node, and the name of the object that calls the member function, `businessContact`, is a child node of the function's name. The nodes provide detailed information, such as each node's ID and the class, `Contact`, of the object. There is one other child node of the `CallExpr` node, which is the argument of the function call. The node states the name of the argument, `cin`, the class, `istream`, and the variable ID.

Leveraging the AST to analyze the code is advantageous because:

- The AST provides ample information for each node, including type and class without recourse to the source code.
- The tree structure allows for easy parsing (again, as compared to the source code).
- The AST allows for dynamic dispatch by displaying dynamic class types for iterators as simply class identifiers (Figure 3), which otherwise might be difficult to extract.

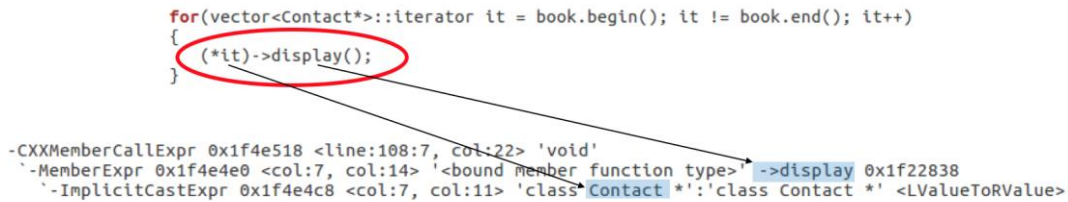

Figure 3: C++ Abstract Syntax Tree, Resolving Dynamic Dispatch

Figure 3 depicts the Clang AST translation of an iterator. The class associated with the member function, `display`, is not of the iterator, but of the class of the object to which the iterator points. This allows the resulting call graph to display the correct class associated with the function.

An *Island Grammar* (*IG*) (Moonen, 2002) (Synytskyy, Cordy, and Dean, 2003) is a Context-Free Grammar (CFG) that catches a specific lexicon (*island*) and ignores all the rest (*water*). While other CFGs operate by either accepting a word that belongs to the Context-Free Language (CFL) or rejecting a word that does not, an IG accepts all the words that belong to the language and *ignores* the rest. According to (Moonen, 2002), an Island Grammar can be defined as follows:

> **Definition 2.2** *Given a language $L_0$, a context free grammar $G = (V, \Sigma, P, S)$ such that $L(G) = L_0$ and a set of constructs of interest $I \subset \Sigma^*$ such that $\forall i \in I : \exists s_1, s_2 \in \Sigma^* : s_1 \, i \, s_2 \in L(G)$. An island grammar $G_I = (V_I, \Sigma_I, P_I, S_I)$ for $L_0$ has the following properties:*
> 1. *$L(G) \subset L(G_I)$   $G_I$ generates an extension of $L(G)$.*
> 2. *$\forall i \in I : \exists v \in V_I : v \xrightarrow{*} i$*
>    *$\exists s_3, s_4 \in \Sigma^* : s_3 \, i \, s_4 \notin L(G) \wedge s_3 \, i \, s_4 \in L(G_I)$*
>    *$G_I$ can recognize constructs of interest from $I$ in at least one sentence that is not recognized by $G$.*
> 3. *$K(G) > K(G_I)$   $G$ has higher complexity than $G_I$.[2]*

An IG is a good choice for parsing an AST because of the following:

- IGs are fast, because they will only search for specific words in an AST (and ignore the rest – and ASTs can be large).
- IGs are robust because the rules only apply to a small portion of the AST and are unaffected by the context or syntax changes in the remaining AST.

In (Moonen, 2002), the Syntax Definition Formalism (SDF) (Heering, Henriks, Klint, and Rekers, 1989) is used to present IGs; here we chose to use the common tools *Bison* and *Flex*. *Flex* defines the specific lexicon that the grammar accepts, such as certain keywords in the AST, e.g., `CallExpr`. *Flex* also uses regular expressions to catch groups of words. For example:

$$\text{"\""(\\.|[^"\\])*\""}$$

will catch all the strings in the textual AST. When these keywords or defined groups of words are caught and recognized by the program, *Flex* sends a token identifying the word to *Bison*. The *Bison* program defines the CFG by a collection of rules using terminal and non-terminal language symbols. All of these symbols are defined in *Flex* as the lexicon. Once a word is accepted in *Bison*, C++ is used for further evaluation to create an output CSV format data file with all the function calls and their related information.

```
"CallExpr"              { return CALL; }           (1)
"DeclRefExpr"           { return ARGUMENT; }       (2)
[a-zA-Z][a-zA-Z0-9]     { return WORD; }           (3)
"\""(\\.|[^"\\])*"\""   { return STRING; }         (4)
.                       { // everything else - water }  (5)
```

Figure 4: Flex Lexical Syntax

Figure 4 depicts a simplified version in *Flex* of the lexical syntax of the C/C++ AST for extracting function calls. In line 1, `CallExpr` is the keyword used in the AST to denote that the block pertains to a function call. When this line is found in the AST, *Flex* returns the token CALL. Line 2 is similar, but the keyword `DeclRefExpr` is in connection with an argument of the function call.

Lines 3 and 4 utilize regular expression in lieu of a keyword. Anytime *Flex* comes across a word (or group of words in the case of a string) that fits the regular expression, the appropriate token is returned.

In line 5, the period (.) represents everything else that *Flex* could potentially come across that was not specifically defined. This is considered *water*. Because there is no code associated with the period, *Flex* is instructed to do nothing, or ignore the lexicon. Thus, *Flex* only acts when a line in the AST could potentially be useful in obtaining details about function calls.

```
input:                                                  (1)
    input water | input land | ε                        (2)
water:                                                  (3)
    WORD | STRING                                       (4)
land:                                                   (5)
    CALL WORD          { add_call_name($2); }           (6)
    | ARGUMENT WORD    { add_arg($2); }                 (7)
    | ARGUMENT STRING  { add_arg($2); }                 (8)
```

Figure 5: Bison Context-Free Syntax

Figure 5 depicts a simplified version of the *Bison* rules for the CFL for finding function calls on the C/C++ AST. Line 1 is the start nonterminal symbol, **input**. Every word in the AST starts its analysis at **input**. The recursive rule for **input** is in line 2. **input** is either replaced by the nonterminal symbol **water** or **land**. The terminal epsilon (ε) ends the recursion.

The rules for the nonterminal symbol **water** are stated in lines 3 and 4. **Water** can either be replaced by the terminals WORD or STRING, which are defined in Figure 4 on lines 3 and 4 respectively. The nonterminal **water** ensures that any WORD or STRING that is found in the AST outside of the specific lines for function calls are ignored.

The rules for the nonterminal symbol **land** are found in lines 5 through 8. The first terminal option, on line 6, states that every time the token CALL is found in the AST, which is represented by the keyword `CallExpr` defined in Figure 4 line 1, there will be a token WORD, also defined in Figure 4, immediately afterwards. The C++ code associated with this rule then sends the string value of WORD, (in the variable $2), to the PAIGE function `add_call_name`.

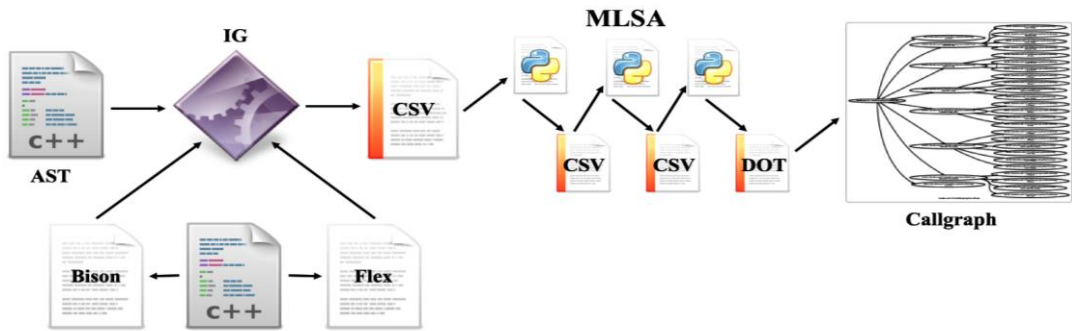

Figure 6: MLSA Architecture with Island Grammar

The second option, on line 7, is for when the AST comes across an argument block in which the argument is a variable. The token ARGUMENT, which is defined in Figure 4 line 2, begins the line and is immediately followed by a WORD token, which is the variable name of the argument. The value of WORD is then sent to the function `add_arg`.

The last option, on line 8, is chosen when an ARGUMENT token is followed directly by a STRING token, defined in Figure 4 line 4. Similar to line 7, the value of STRING is sent to `add_arg` on line 8.

## 4 CALL GRAPH ANALYSIS

The PAIGE filter is the first step in the production of a multilingual call graph. These call graphs contain the name of the function called as well as the arguments in the call and the class (if the function is a member function). In the MLSA architecture, illustrated in Figure 1, the PAIGE filter is at the monolingual level, where programs are analyzed separately depending on the language. An IG for function calls is designed for each language to be processed. Our examples are limited to C/C++ in this paper; we have built IGs also for Python and JavaScript. The IG for C/C++ function calls was designed to analyze the *Clang* AST; it was written specifically to fit the C/C++ AST format and keywords. The IG analyzes the AST and prints the results in a CSV format output data file, which can then be analyzed in MLSA's multilingual level, as shown in Figure 6. The multilingual level incorporates all programs, regardless of language, to create a multilingual call graph in a CSV format output file. The open source *Graphviz* program is used to generate a DOT format file containing a visual representation of the call graph; Figure 7 shows an example call graph generated by PAIGE.

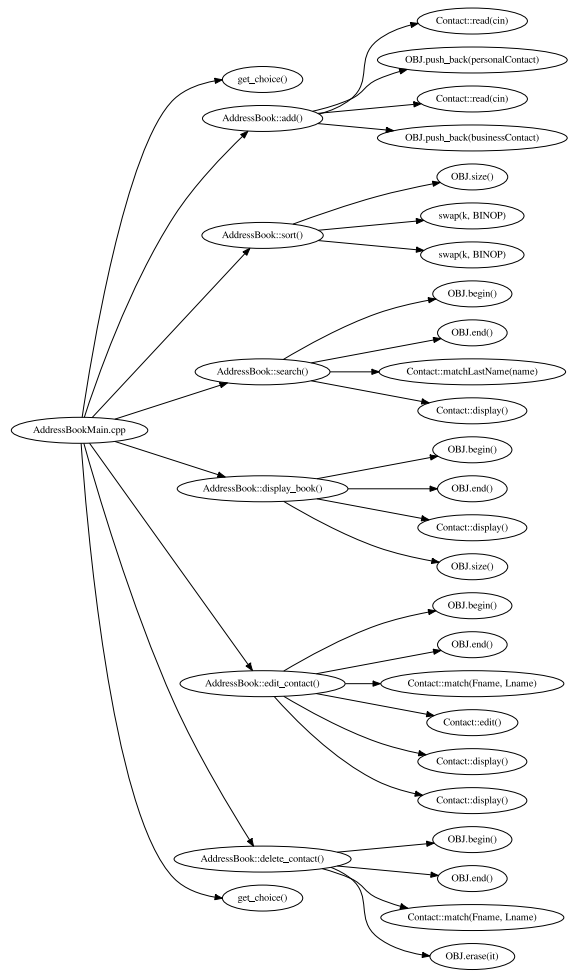

Figure 7: Example C++ Call Graph

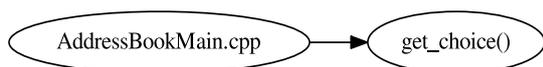

Figure 8: Straightforward Call Graph

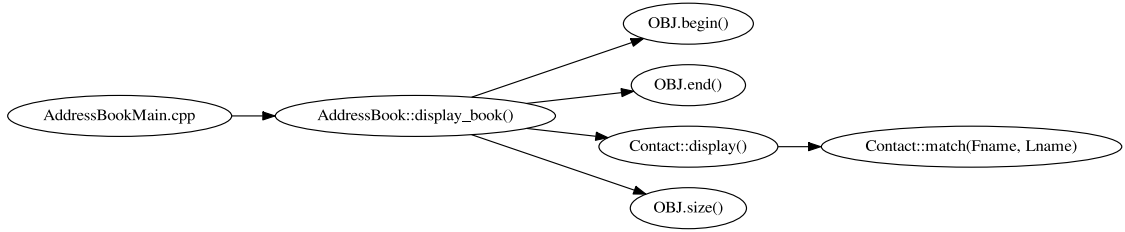

Figure 9: More Involved Case Call Graph

Figure 8 is a simple call graph generated by MLSA. `get_choice` is a function defined in the body of the C++ program, `AddressBookMain.cpp`. The function is called by the main function in the program. `get_choice` also has no arguments passed to it. These three factors make the function call quite simple to parse and display. The IG need only retrieve the name of the function called and the scope of where that function was called (for this example the scope would be the main function in `AddressBookMain.cpp`).

Figure 9 represents a call graph which is a bit trickier to parse. In this example, there are functions called which are defined in three different C++ programs – `AddressBookMain.cpp`, `AddressBook.cpp` and `Contact.cpp`. `AddressBookMain.cpp` can determine that it is calling `display_book` from `AddressBook.cpp`, but it cannot determine which functions are called inside `display_book`'s definition. Similarly, `AddressBook.cpp` can determine that `display_book` calls `display` from `Contact.cpp`, but it does not know whether or not `display` calls any functions. PAIGE must not only track which function is being called, but also which program contains the definition of that function so that the function calls can be connected later on in the call graph.

Another complication in this example is that both `display_book` and `display` are class member functions. The IG must determine that (1) the function is a member function, and (2) of which class the function is a member. If the function is a member of a class defined in the program, the call graph displays the function name as `Class::function`. If the function is a member of a class defined in a library, the function is displayed as `OBJ.function` to distinguish the two.

The third complication cannot specifically be detected in the function. Inside the `Contact` member function `display`, `match` is called via the implied pointer *this*, which refers to the object which called the `display` function. PAIGE determines the class of the *this* pointer, which is displayed differently in the AST, and presents the function call as `Contact::match` in the call graph.

The function `Contact::match` also has two arguments, `Lname` and `Fname`. PAIGE must determine what type of arguments they are – for this example the two arguments are variables – and the names of the arguments.

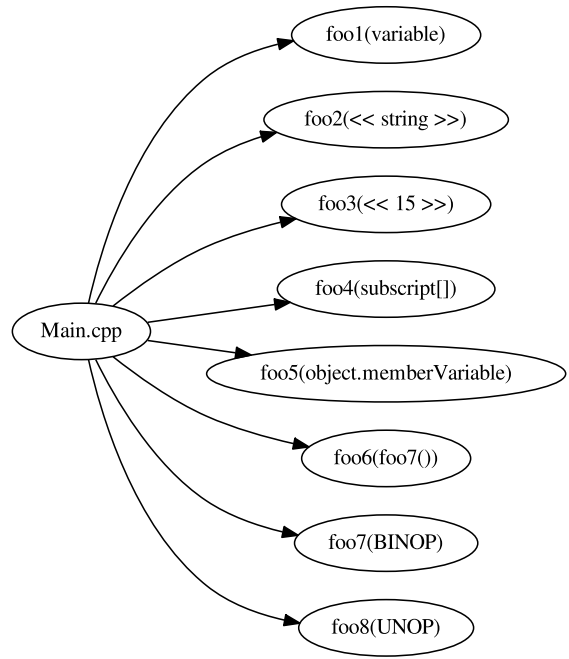

Figure 10: Call graph with Arguments

The other types of arguments are displayed in Figure 10. In descending order: (1) variable, (2) string, (3) integer or double (4) subscript, such as vector[0], (5) member variable, with the name of the object, (6) the returned value of another function call, displayed as the function call, (7) a binary operation, such as 2+2, (8) a unary operation, such as k++.

It was decided early on to have PAIGE produce a tree structured call graph rather than a general graph. A tree is arguably a simpler construct for a designer to understand. Since PAIGE shows call arguments, different calls to the same function are in fact different branches anyway, and any recursive calls are simply marked as recursive leaf nodes.

## 5 EXPERIMENTATION AND RESULTS

### 5.1 Doxygen Comparison

To test the performance of the PAIGE approach, the call graphs generated by PAIGE were manually compared to those of the *Doxygen* automated documentation system. *Doxygen* was developed to automatically generate documentation from source code, including call graphs. It uses a simple parser for this and hence is lightweight in a similar sense to PAIGE and is a good comparison. However, it deals with the source code directly, whereas PAIGE looks at the AST.

One hundred C/C++ programs were collected for the experiment from online source repositories. Programs were selected only if they produced a *Doxygen* graph – that is, they needed to call at least one function in the main function which was defined in the program. This is because *Doxygen* does not show library function calls in its call graphs. Programs that included classes were preferred.

Both *Doxygen* and PAIGE were run on each program, and the DOT files (visual representation of the call graph, as in e.g. Figure 7) for each call graph were manually compared. Because *Doxygen* produces a graph and PAIGE produces a tree, only one node for each function called was kept for comparison. A better experiment would have been to compare each individual function call to make sure all were caught, but since Doxygen only shows one node per function, this comparison was impossible.

Six features were measured for each call graph:

1. the total number of function calls caught by each approach,
2. the number of member function calls caught,
3. the number of regular function calls caught,
4. the number of library function calls caught,
5. the number of arguments caught, and
6. the number of programs where recursion is correctly listed.

Figure 11 shows the results of these six measurements for the 100-program codebase for the call graphs generated by *Doxygen* and those by PAIGE. As can be seen, PAIGE collects twice as many function calls as *Doxygen*, but about ¾ of those are library function calls that *Doxygen* does not show. The remaining ¼ of calls that PAIGE catches that *Doxygen* does not are member functions.

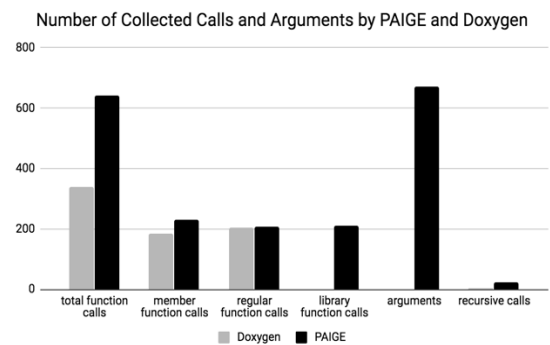

Figure 11: Doxygen and PAIGE Comparison Results

The main reason that MLSA catches about 45 more member functions than *Doxygen* is that PAIGE catches functions that are called with the implied *this* pointer. The *this* pointer is used in member function definitions and refers to the class object. The *this* pointer can either be used when calling a member function of the same class or when accessing member variables. The pointer can be implied, meaning that *this* is not specifically written but will still be invoked. Because the *this* pointer information is explicit in the AST, PAIGE catches both member functions called inside other member functions as well as functions of another class called inside member functions where the object of the other class is a member variable of the first class. *Doxygen* catches neither as it does not parse the *this* pointer. An example of how the implied *this* pointer looks in the C++ AST is presented in Figure 12.

Figure 12: C++ AST with *this* Pointer

In Figure 12, `m_drawingAPI` is a `DrawingAPI`-type member variable of class `CircleShape`. In the AST, not only is the class of `m_drawingAPI` given, but the class of the *this* pointer (`CircleShape`) is also given. Therefore, PAIGE is able to connect the function `drawCircle` to the function definition in the `DrawingAPI` class definition.

There are two principal reasons for lapses in PAIGE's ability to catch functions that *Doxygen* catches. The first principal reason is the state of development of the *Clang* AST: it does not currently accept template classes, for example, and does not include this code in its AST, so any member functions of a template class will not be displayed in a PAIGE call graph. Also, the *Clang* AST views static functions as regular functions instead of member functions and therefore does not include the function's class. PAIGE does catch this function, but displays it as a regular function (and therefore cannot include any more function calls inside that function's definition).

The second principal reason that PAIGE may catch fewer function calls than *Doxygen* relates only to the current PAIGE IG: information is available in the *Clang* AST, but has not yet been addressed in the PAIGE IG. For example, using namespaces in class definitions: The namespace and class are included in the function call, but only the class name is included in the class definition right now, causing some functions not to show up in the final call graph. But with some addition to the IG code, this problem could be easily addressed. Another issue is the use of function pointers: *Doxygen* catches the name of the function, while PAIGE only catches the function and displays a warning that the name of the function cannot be caught. The function pointer is displayed differently in the C++ AST than a regular function call, so additional work needs to be done to catch the name of the function. Handling polymorphism also falls into this class.

One difference between PAIGE and *Doxygen* is that PAIGE displays library function calls. This is necessary because MLSA is meant to process programs written in multiple computer languages and display how they interact. Many of these languages interact through library functions; for example, `Python.h` allows python code to be called in a C/C++ program, so library functions are *key* for catching interoperability.

PAIGE also catches all the arguments in call functions – recall that the PAIGE call graph is a tree and not a graph – for the same purpose that it catches library calls. These library functions that allow interoperability take in arguments that indicate what cross-language code is being called, giving for example the name of a program or the actual code itself.

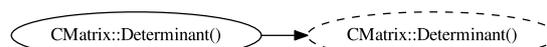

Figure 13: PAIGE Recursive Function

Lastly, call graphs were compared to determine whether recursion is displayed correctly. PAIGE distinguishes a recursive function call by drawing it as a leaf node and bordering the node in a dotted line, as seen in Figure 13. *Doxygen* only caught recursive functions in two of the 22 programs that contained recursive functions. Catching recursion is clearly important for analyzing a program.

## 5.2 Clang AST for Objective-C

*Clang* can process Objective-C source code as well as C/C++ code. In theory, PAIGE should be able to generate call graphs for Objective-C from the *Clang* AST. PAIGE was tested on five Objective-C programs. In fact, MLSA did catch all function calls written in the C/C++-format that used the keyword `CallEpr`, which some programs included. However, it could not catch the majority of Objective-C function definitions and function calls, as they were displayed in the AST with Objective-C-specific keywords – `ObjCMethodDecl` and `ObjCMessageExpr` respectively; nonetheless, the IG parsing did not crash either. Objective-C call syntax can be easily included into the IG by just adding a few keywords.

## 5.3 Time Comparison and Robustness

To evaluate the time performance of the IG approach, PAIGE was tested against the original call graph approach used in MLSA – ad-hoc string parsing with Python. Fifty-five of the 100 programs were chosen on which to test the speed of each approach's execution. The programs were chosen based on the criteria that they could in theory be correctly processed into a call graph by the ad-hoc software (which did not handle member functions among other things). In the experiment, each program was analyzed with both the ad-hoc approach and the PAIGE approach, the time for each program to be analyzed was recorded, and the results are shown in Figure 14.

As can be seen in Figure 14, PAIGE outperforms the ad-hoc approach by a factor of about 1:3. PAIGE averaged a time per program of about 0.15 seconds, whereas the ad-hoc approach averaged roughly 0.62 seconds per program.

Furthermore, while running the ad-hoc approach on the 55 programs, the program crashed on 5 of the runs. The reason for the crash has to do with an incorrect parsing of the C/C++ AST by the ad-hoc approach. The approach crashed while searching for a specific line in the AST. Instead of parsing the AST by lines and positions, the PAIGE approach parses the AST by keywords. Not only is this approach faster, but it also eliminates potential crashes.

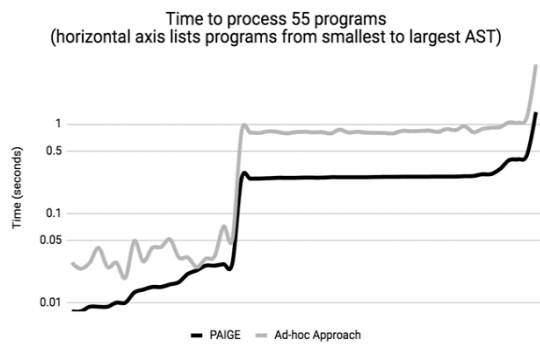

Figure 14: MLSA and PAIGE Time Comparison

# 6 DISCUSSION

This paper has presented a novel lightweight tool for analyzing multilingual codebases, a topic of increasing importance to large software companies. Within the context of our ongoing development of MLSA – an architecture for Multilingual Software Analysis – we have presented a lightweight and modular approach to monolingual call graph generation which can nonetheless provide sufficient information about function calls and arguments to allow interoperability API processing and the construction of a single multilingual call graph from a set of monolingual call graphs (Lyons, Bogar, and Baird, 2017). The input to the call graph generator PAIGE is a monolingual AST. The paper has focused on a C/C++ monolingual generator using the *Clang* AST as input, but we have constructed similar tools for JavaScript and Python. The *Clang* AST is very rich in detail, allowing PAIGE to address (partially) more difficult issues such as dynamic dispatch and iterators. An Island Grammars (Moonen, 2002) was used to specify monolingual call syntax in a robust (with respect to changes in the AST syntax) and lightweight fashion. Performance results were presented to show that PAIGE performed better than *Doxygen*, another more lightweight call graph generator, performed faster than a call graph generator that processed the AST text in an ad-hoc fashion, and showed robustness that the ad-hoc version could not follow.

Many solutions to static call graph extraction exist (see (Hoogendorp, 2010) for a nice summary). Techniques for finding the target set of a call expression such as CHA (Dean, Grove, and Chambers, 1995), RTA (Bacon and Sweeney, 1996), XTA (Tipp and Palsberg, 2000) are obliged to do extensive reasoning about the program. By leveraging the AST, PAIGE offloads this reasoning to the compiler. This is a sensible division of responsibilities; compilers will improve or change in functionality to reflect the needs of the programming community, and PAIGE can benefit from this trend.

PAIGE is not alone in using an AST for input; The *Eclipse* IDE make the compiler AST available for static analysis applications. However, many approaches go the route of *Doxygen* and write their own parser (e.g., Rigi (Kienle and Müller, 2010)) – resulting in a potentially poorer performance than using a full compiler, as seen here for *Doxygen*. PAIGE has a specific parser – a very simple one thanks to the use of Island Grammars – but it applies this to the AST rather than the source code, and hence is a good balance of lightweight and yet strong performance. The MLSA software repository containing the PAIGE software can be made available by emailing the paper authors.

## ACKNOWLEDGEMENTS

The authors acknowledge the contributions of Bruno Vieira and Sunand Raghupathi in developing the results in this paper. The authors are partially supported by grant DL-47359-15016 from Bloomberg L.P.